\documentclass[11pt]{article}
\newcommand{\BE}{\begin{equation}}
\newcommand{\EE}{\end{equation}}
\newcommand{\BA}{\begin{eqnarray}}
\newcommand{\EA}{\end{eqnarray}}

\begin{document}
\begin{titlepage}

\vspace*{1mm}
\begin{center}

            {\LARGE{\bf A weak, attractive, long-range force \\
                        in Higgs condensates }}

\vspace*{14mm}
{\Large  M. Consoli }
\vspace*{4mm}\\
{\large
Istituto Nazionale di Fisica Nucleare, Sezione di Catania \\
Corso Italia 57, 95129 Catania, Italy}
\end{center}
\begin{center}
{\bf Abstract}
\end{center}

Due to the peculiar nature of the underlying medium, density fluctuations
in  a `Higgs condensate' are predicted to propagate
for infinitely long wavelengths 
with a group velocity $c_s\to \infty $. On the other hand, 
for any large but finite $c_s$ there is a weak, attractive $1/r$
potential of strength ${{1}\over{c^2_s}}$ 
and the energy spectrum deviates from the purely 
massive form $\sqrt{ {\bf{p}}^2 + M^2_h}$ at momenta smaller than
$\delta\sim {{M_h}\over{c_s}}$. Physically,
the length scale $\delta^{-1}$ corresponds to the mean free-path for the elementary
constituents in the condensate and would naturally be placed in the 
millimeter range.

\vskip 35 pt
\end{titlepage}

{\bf 1.}~~In this Letter, I shall discuss some phenomenological aspects of the 
ground state of spontaneously broken theories: the `Higgs condensate'. The name 
itself (as for the closely related gluon, chiral,..condensates) indicates that, 
in our view, this represents a kind of medium 
made up by the physical condensation process of some
elementary quanta. If this were true, 
such a vacuum should support long-wavelength
density fluctuations. In fact, the existence of density fluctuations
in {\it any known medium}
is a basic experimental fact depending on the coherent response of
the elementary constituents to disturbances whose wavelength is much larger
than their mean free path. This leads to an universal description, the
 `hydrodynamical regime', that does not depend on the details of the
underlying molecular dynamics. 
By accepting this argument, 
and quite independently of the Goldstone phenomenon, the energy spectrum 
of a Higgs condensate should terminate with 
an `acoustic branch', say $\tilde{E}({\bf{p}})=c_s|{\bf{p}}|$
for ${\bf{p}} \to 0$, 
as for the propagation of sound waves in ordinary media. 

Some arguments suggest that, indeed, the vacuum of
a `pro forma' Lorentz-invariant quantum field theory may be
such kind of medium. 
For instance, a fundamental phenomenon as
the macroscopic occupation of the same quantum state 
(say ${\bf{p}}=0$ in some frame) may 
represent the operative construction of 
a `quantum aether' \cite{dirac,volovik}. This would be
quite distinct from the 
aether of classical physics, considered a truly preferred
reference frame and whose constituents were
assumed to follow definite space-time trajectories.
 However, it would also be different
from the empty space-time of special relativity, assumed at the base of
axiomatic quantum field theory to deduce the exact Lorentz-covariance of
the energy spectrum.

In addition, one should take into account the 
approximate nature of locality in
cutoff-dependent quantum field theories. 
In this picture, the elementary quanta are treated as `hard spheres', 
as for the molecules of ordinary matter. Thus, the notion of the
vacuum as a `condensate' acquires an intuitive physical meaning. 
For the same reason, however, the simple idea that all deviations from 
Lorentz-covariance take place at the cutoff scale may be incorrect.
In particular, non-perturbative vacuum condensation may
give rise to a hierarchy of scales
such that the region of Lorentz-covariance is sandwiched {\it both}
in the high- and low-energy region.

In fact, in general, 
an ultraviolet cutoff induces vacuum-dependent
{\it reentrant violations of
special relativity in the low-energy corner} \cite{volo2}. In the simplest
possible case, these extend over a small shell of momenta, say 
$|{\bf{p}}|< \delta$, where the energy spectrum $\tilde{E}({\bf{p}})$ deviates
from a Lorentz-covariant form. Since 
Lorentz-covariance becomes an exact symmetry in the local limit,
for very large $\Lambda$, the scale $\delta$ is naturally
infinitesimal in units of the energy scale associated
with the Lorentz-covariant part of the energy spectrum, say $M$. 
By introducing dimensionless quantities, this means 
$\epsilon\equiv {{\delta}\over{M}} \to 0$ when
$t \equiv {{\Lambda}\over{M}} \to \infty$ so that
the continuum limit can equivalently be defined either as $t \to \infty$ or
$\epsilon \to 0$. Notice that, formally, 
${\cal O}({{\delta}\over{M}})$ vacuum-dependent corrections 
represent ${\cal O}({{M}\over{\Lambda}})$ effects which are always
neglected when discussing \cite{nielsen} how
Lorentz-covariance emerges at scales much smaller than the ultraviolet 
cutoff. In this sense, Lorentz-covariance is formally valid in the local
limit but, for finite $\Lambda$, there are infinitesimal deviations
in an infinitesimal region of momenta that cannot be
understood without exploring the physical properties of the vacuum. 

For instance, for large but finite $t$, does the group velocity
$c_s={{d\tilde{E}}\over{ d|{\bf{p}}| }}$ remain below unity (in units of
the light velocity $c$) when $|{\bf{p}}| \to 0 $ ?
Of course, a group velocity $c_s >1$ 
does not correspond to any `classical'
velocity being faster than light. Moreover, 
when $t \to \infty$, the relevant wavelengths are ${\cal O}(1/\epsilon)$  
and thus infinitely long
on the physical length scale defined by $1/M$. As a consequence, in a strict
continuum limit there would be no way to form the sharp wave
fronts needed to transfer any type of
information. However, what about the cutoff theory ? For instance, 
if we just play with the
numbers, and choose a mass unit $M={\cal O}(10^2)$ GeV with
values of $t$ as large as
${\cal O} (10^{16})$, we find conceivable to observe deviations
from Lorentz-covariance 
at scales ${\cal O}(10^{-33})$ cm.
On the other hand, if such deviations
were {\it reentrant} with an $\epsilon \sim 1/t$, what about vacuum fluctuations
whose wavelengths were larger than a few millimeters ?
Should we consider them as 
`infinitely' long ? In this sense, a non-trivial structure of the vacuum
raises delicate issues for which no general answers exist `a priori'.

\vskip 10pt

{ \bf 2.}~~After this preliminary discussion, 
our analysis will start after having understood that, 
quite independently of the Goldstone phenomenon, there is
a gap-less mode of the singlet  Higgs field
in the spontaneously broken phase of $\lambda \Phi^4$ 
theories. 

The presence of such a gap-less mode reflects the quantum 
nature of the scalar condensate that cannot be treated as a purely 
classical c-number field. In fact, either considering the 
re-summation of the one-particle reducible
zero-momentum tadpole graphs \cite{singlet} in a given background field 
or performing explicitely the last functional integration over the strength
of the zero-momentum mode of the singlet Higgs field
\cite{legendre}, one finds {\it two} different solutions
for the inverse zero-4-momentum propagator in 
the  spontaneously broken phase of
a (one component) $\lambda\Phi^4$ theory:
a)~$G^{-1}_a(p_\mu=0)=M^2_h$ and b)~$G^{-1}_b(p_\mu=0)=0$. 
For the convenience of the reader, we shall briefly repeat the
main argument of ref.\cite{legendre}.

When discussing spontaneous symmetry breaking, the starting point 
is the separation of the scalar field into a constant background
and a shifted fluctuation field, namely 
\BE
\label{shift}
          \Phi(x)= \phi + h(x)
\EE
In order Eq.(\ref{shift}) to be unambiguous, 
$\phi$ denotes the spatial average in a large 4-volume $\Omega$ 
\BE
\label{average}
          \phi= {{1}\over{\Omega}}\int d^4x~ \Phi(x)
\EE
and the limit $\Omega \to \infty$ has to be taken at the end.

In this way, the full functional measure can be expressed as
\BE
\label{measure}
                \int[d\Phi(x)]...=\int^{+\infty}_{-\infty}d\phi\int[dh(x)]...
\EE
and the functional integration on the r.h.s. of Eq.(\ref{measure}) 
is over all quantum modes for $p_\mu\neq 0$. 

After integrating 
out all non-zero quantum modes, 
the generating functional in the presence of a space-time constant 
source $J$ is given by
\BE
\label{zetaj}
      Z(J)= \int^{+ \infty}_{-\infty} d\phi~ \exp [-\Omega
(V_{\rm NC}(\phi) - J\phi)]
\EE 
and
$V_{\rm NC}(\phi)$ denotes the usual non-convex (`NC') effective potential
obtained order by order 
in the loop expansion. Finally, by introducing the generating functional for
connected Green's functions $w(J)$ through
\BE
\label{log}
\Omega~ w(J)=\ln {{Z(J)}\over{Z(0)}}
\EE
one can compute the field expectation value
\BE
\label{phij}
\varphi(J)={{dw}\over{dJ}}
\EE
and the zero-momentum propagator
\BE
\label{GJ}
G_J(p_\mu=0)={{d^2w}\over{dJ^2}}
\EE
In this framework, 
spontaneous symmetry breaking corresponds to non-zero values of 
Eq.(\ref{phij}) in the double limit $J \to \pm 0$ and 
$\Omega \to \infty$. 

Now, by denoting $\pm v$ the absolute minima of 
$V_{\rm NC}$ and 
$M^2_h=V''_{\rm NC}$ its quadratic shape there, one usually assumes 
\BE
\label{ssb}
\lim_{\Omega \to \infty} \lim_{J \to \pm 0} \varphi(J) = \pm v
\EE
and
\BE
\label{GA}
\lim_{\Omega \to \infty} \lim_{J \to \pm 0} G_J(p_\mu=0)={{1}\over{M^2_h}}
\EE
In this case, the excitations in the broken phase 
would be massive particles (the conventional Higgs bosons)
whose mass $M_h$ is determined by the positive 
curvature of $V_{\rm NC}$ at its absolute minima.

The main result of \cite{legendre} is that this conclusion
is not true. In fact, at $\varphi=\pm v$, 
besides the value ${{1}\over{M^2_h}}$, one also finds 
\BE
\label{GB}
\lim_{\Omega \to \infty} \lim_{J \to \pm 0} G_J(p_\mu=0)=+\infty
\EE
a result that has no counterpart in perturbation theory. 
As discussed in ref.\cite{legendre}, 
the existence of such divergent behaviour admits a
simple geometric interpretation in terms of 
the Legendre transform of $w(J)$. Differently from 
$V_{\rm NC}$, this other definition of effective potential does not provide 
an infinitely differentiable
function in the presence of spontaneous symmetry breaking \cite{syma}.
Therefore, its 
left- and right- second derivatives at $\varphi=\pm v$ do {\it not} coincide.
In this sense, the existence of a singular zero-4-momentum propagator in the
broken phase is
a genuine quantum-field theoretical effect, quite independent of any 
physical interpretation. This will be discussed below.
\vskip 10 pt

{\bf 3.}~~`A priori', the existence of 
two possible values for the zero-4-momentum propagator implies
two possible types of excitations with the same quantum numbers but
different energies when the 3-momentum
${\bf{p}} \to 0$: 
a massive one, with $\tilde{E}_a({\bf{p}}) \to M_h$, and a gap-less one with 
$\tilde{E}_b({\bf{p}}) \to 0$. However, the latter dominates the 
exponential decay $\sim e^{-\tilde{E}_b({\bf{p}})T}$  of the
connected euclidean correlator for
${\bf{p}} \to 0$. In this sense, 
the massive excitation is unphysical in the infrared region.
Therefore, differently from the simplest perturbative indications, 
in a (one-component) spontaneously broken
$\lambda\Phi^4$ theory there is no energy-gap associated
with the `Higgs mass'  $M_h$, 
as it would be for a genuine massive single-particle 
spectrum where the relation
\BE
\label{Ea}
\tilde{E}_a({\bf{p}})=\sqrt{ {\bf{p}}^2 + M^2_h} 
\EE
remains true for ${\bf{p}} \to 0$. Rather, the infrared region
is dominated by gap-less collective excitations
\BE
\label{Eb}
\tilde{E}_b({\bf{p}}) \equiv c_s |{\bf{p}}| 
\EE
depending on an unknown parameter $c_s$ that controls the slope of the spectrum
for  ${\bf{p}} \to 0$ and represents the `sound velocity' 
for the density fluctuations of the scalar condensate. 
The massive branch, however, can become important at higher momenta.
To understand this point, let us
explore the analogy with superfluid $^4$He. This analogy
is based on the observation
\cite{mech} that, as for the 
interatomic $^4$He-$^4$He potential, the low-energy limit of cutoff
$\lambda\Phi^4$ is also
a theory of quanta with a short-range repulsive core and
a long-range attractive tail \cite{note0}.
 
The essential point is that, for superfluid $^4$He, 
the existence of two types of
excitations  was first deduced theoretically by Landau
on the base of very general arguments \cite{landau}. According to this original
idea, there are phonons with energy
$E_{\rm ph}({\bf{p}}) = v_s |{\bf{p}}|$ and
rotons with energy
$E_{\rm rot}({\bf{p}})= \Delta + {{ {\bf{p}}^2 }\over{2\mu}}$. 
Only {\it later},
it was experimentally discovered that
there is a single energy spectrum $E({\bf{p}})$ which is
made up by a continuous matching of these two
different parts. This unique spectrum agrees with the phonon branch
for ${\bf{p}} \to 0$ and agrees with the roton branch 
at higher momenta.

Although these results are well established, 
all details of the energy spectrum of superfluid $^4$He 
in the matching region are not yet completely  
understood \cite{helium}. For this reason, in our case of
the spontaneously broken phase, 
we shall just extract the main conclusion: 
the existence of a single energy spectrum 
$\tilde{E}({\bf{p}})$ that tends to
$\tilde{E}_b({\bf{p}})$ for ${\bf{p}} \to 0$ and approaches
$\tilde{E}_a({\bf{p}})$ for larger values of
$|{\bf{p}}|$. 

However, for $\Delta=\mu\equiv M_h$, 
the matching between Eqs.(\ref{Ea}) and (\ref{Eb}) 
is only possible for $c_s>1$, 
in units of the light velocity $c$ (this can easily be checked 
looking for a possible intersection between
Eqs.(\ref{Ea}) and (\ref{Eb})).
Only in this case, 
for sufficiently high momenta where
$\tilde{E}_a({\bf{p}})< \tilde{E}_b({\bf{p}})$, the gap-less collective modes
will become unphysical and the lowest excitations of the vacuum will
correspond to the familiar, massive Higgs boson.

Independently of this argument, 
the idea that $c_s$ is actually {\it infinitely} 
larger than the light velocity
is also supported by a semi-classical
argument due to Stevenson \cite{seminar} that we shall briefly report.
Stevenson's argument starts from a perfect-fluid treatment of the Higgs
condensate. In this approximation, 
energy-momentum conservation is equivalent to wave propagation with a
squared velocity given by ( $c$ is the light velocity)
\BE
\label{cs}
        c^2_s= c^2 ({{\partial {\cal P}}\over{\partial {\cal E} }})
\EE
where ${\cal P}$ is the pressure and ${\cal E}$ the
energy density. Introducing the condensate density $n$, and using
the energy-pressure relation
\BE
\label{zerot}
 {\cal P}= -{\cal E} + n {{\partial {\cal E} }\over{\partial n}}
\EE
we obtain
\BE
\label{nn1}
c^2_s=
c^2({{\partial {\cal P}}\over{\partial n}})
({{\partial {\cal E} }\over{\partial n}})^{-1}=
c^2  (n{{ \partial^2 {\cal E} }\over{\partial n^2}})
({{\partial {\cal E} }\over{\partial n}})^{-1} 
\EE
For a non-relativistic Bose condensate of neutral particles with mass
$m$ and scattering length $a$, where ${{na\hbar^2}\over{m^2c^2}} \ll 1$
(in this case we explicitely introduce $\hbar$ and $c$) one finds
\BE
\label{nonrel}
{\cal E}= n m c^2 + n^2 {{2\pi a \hbar^2}\over{m}} 
\EE
so that 
\BE
\label{nn2}
c^2_s={{4\pi n a \hbar^2}\over{m^2}}
\EE
which is the well known result for the sound velocity in a  
dilute hard sphere Bose gas \cite{huang}.

 On the other hand, in a fully relativistic case, the additional terms
in Eq.(\ref{nonrel}) are such that the scalar condensate 
is spontaneously generated from the `empty' vacuum where $n=0$ for
that particular equilibrium density where \cite{mech}
\BE
\label{vacuum}
{{\partial {\cal E} }\over{\partial n}} = 0
\EE
Therefore, in this approximation, approaching the equilibrium density
one finds
\BE
\label{infinity}
           c^2_s \to \infty
\EE
thus implying that long-wavelength
density fluctuations would propagate instantaneously in the 
spontaneously broken vacuum. 

As Stevenson points out \cite{seminar}, 
Eq.(\ref{infinity}) neglects all possible corrections to the perfect-fluid
approximation. These require to introduce the effects
of a mean free path $R_{\rm mfp} $
for the elementary constituents. Due to its finite value, 
in fact, sound waves will stop to propagate 
at a typical momentum
$|{\bf{p}}| \sim \delta \equiv {{1}\over{R_{\rm mfp} }}$ \cite{seminar}. 
At this value, the collective modes Eq.(\ref{Eb})
become unphysical and for higher momenta the 
spectrum Eq.(\ref{Ea}) applies. The transition corresponds to
\BE
\label{beyond}
\sqrt{ \delta^2 + M^2_h} \sim c_s \delta
\EE
so that
for $c_s \to \infty$, ${{\delta}\over{M_h}} \sim {{1}\over{c_s}} \to 0$.
Therefore, $c_s$ 
represents the inverse of the parameter $\epsilon$ introduced
before to characterize
the continuum limit. For this reason, we shall replace the condition
$\epsilon \to 0$ with $c_s \to \infty$.

As anticipated, in the continuum limit,
superluminal wave propagation is restricted to the region 
${\bf{p}} \to 0$ and , therefore, one cannot use these waves to 
form a sharp wave front. As such, there would be no possibility 
to transfer informations giving rise to
violations of causality \cite{seminar}. These require
a group velocity
${{d\tilde{E}}\over{ d|{\bf{p}}| }} >1 $ at finite $|{\bf{p}}|$ 
that cannot occur due to the change of the energy spectrum from 
Eq.(\ref{Eb}) to Eq.(\ref{Ea}).

In this sense, the perfect-fluid
value $c_s=\infty$ simulates an exact
Lorentz-covariant limit where 
the energy spectrum maintains its massive form 
for ${\bf{p}} \to 0$. Yet, 
this is not entirely true due to the subtleties
associated with
the zero-measure set ${\bf{p}}=0$. This set, in fact, belongs to the range
of Eq.(\ref{Eb}) and therefore
the right $c_s = \infty$ limit is always
$\tilde{E}({\bf{p}}=0)=0$ and 
{\it not} 
$\tilde{E}({\bf{p}}=0)=M_h$. 
Just for this reason, the correct procedure is to 
include both branches of the spectrum in the spectral representation of 
the fluctuation field $h(x)$. It
may be convenient, however, to separate out 
the long-wavelength modes Eq.(\ref{Eb}), say $\tilde{h}(x)$, 
 from the more conventional massive part of Eq.(\ref{Ea}). 
The observable effects due to
$\tilde{h}(x)$ depend crucially 
 on the value of $c_s$ and will be discussed below.

\vskip 10pt
{\bf 4.}~~Let us ignore, for the moment, the previous indication in Eq.(\ref{infinity})
and just explore the phenomenological implications
of long-wavelength modes 
in the spectrum as in Eq.(\ref{Eb}).
Whatever the value of $c_s$, 
these dominate the infrared region so that
a general yukawa coupling of the Higgs field to fermions will give
rise to a long-range {\it attractive} potential
between any pair of fermion masses $m_i$ amd $m_j$ 
\BE
\label{Newton}
            U_{\infty}(r)= - {{1}\over{4\pi c^2_s
\langle \Phi \rangle^2 }}{{m_im_j}\over{r}}
\EE
The above result would have a considerable impact
for the Standard Model if we take the value $\langle\Phi\rangle \sim 246$ GeV
related to the Fermi constant. Unless $c_s$ be an extremely 
large number (in units of $c$) one is faced with strong
long-range forces coupled to the inertial masses of the known elementary 
fermions that have never been observed. 
 Just to have an idea, for $c_s = 1$ 
the long-range interaction between two electrons in
Eq.(\ref{Newton}) is ${\cal O}(10^{33})$ larger than their purely
gravitational attraction. 
On the other hand, invoking a phenomenologically viable strength,  
as if $c_s \langle\Phi\rangle$ were
of the order of the Planck scale, is equivalent to re-obtain 
nearly instantaneous interactions transmitted by the scalar condensate as 
in Eq.(\ref{infinity}). 

Independently of phenomenology, the identification 
$c_s \langle\Phi\rangle=M_{\rm Planck}$ 
is also natural \cite{seminar} noticing that
for $c_s \to \infty$ 
the energy-spectrum becomes Lorentz-covariant (with the exception
of ${\bf{p}}=0$). Therefore, 
in a picture where the `true' dynamical origin
of gravity is searched into long-wavelength deviations from exact
Lorentz-covariance \cite{volo2}, it would be 
natural to relate the limit of a vanishing gravitational strength, 
$M_{\rm Planck}\to \infty$, to the limit of
an exact Lorentz-covariant spectrum, 
$c_s \to \infty$ \cite{note1}.

Returning to more phenomenological aspects, we observe that the potential 
in Eq.(\ref{Newton}) can also be derived as a static limit 
from the effective lagrangian
($\tilde{\sigma}\equiv {{\tilde{h}}\over{\langle\Phi\rangle}} $)
\BE
\label{lagrangian1}
 {\cal L}_{\rm eff} (\tilde{\sigma})=  
{{ \langle\Phi\rangle^2}\over{2}} \tilde{\sigma}
       [ c^2_s \Delta - {{\partial^2 } \over{\partial t^2}}] \tilde{\sigma}
      -  \tilde{\sigma} 
        \sum_f m_f \bar{\psi}_f \psi_f
\EE
where the free part takes into account the peculiar nature of the energy
spectrum Eq.(\ref{Eb}).
Eq.(\ref{lagrangian1}) is useful to represent the effects of $\tilde{\sigma}$ 
over macroscopic scales as required by its
long-range nature. The key-ingredient is the replacement of
        $m \bar{\psi} \psi$ with 
$T^{\mu}_{\mu}(x)$, the trace of the energy-momentum tensor of ordinary matter,
a result embodied into the well known relation
\BE
        \langle f| T^{\mu}_{\mu}| f \rangle=
         m_f \bar{\psi}_f \psi_f
\EE
This relation allows for an intuitive 
transition from the quantum to the classical
theory. In fact, by introducing 
a wave-packet corresponding to a particle of
momentum ${\bf{p}}$ and normalization 
$\int d^3{\bf{x}} \bar{\psi} \psi={{m}\over{E({\bf{p}})} }$
we obtain 
\BE
\label{replace}
        - m 
\int d^4x \bar{\psi} \psi~ =~-m \int ds
\EE
where $ds=dt\sqrt{1-{\bf{v}}^2}$ 
denotes the infinitesimal element of proper time for a classical
particle with velocity ${\bf{v}}$. Therefore, using the relation
\BE
         \sum_n m_n \int ds_n= \int d^4x T^{\mu}_{\mu}(x)
\EE
where
\BE
         T^{\mu}_{\mu}(x) \equiv \sum_n 
{ { E^2_n -   {\bf{p}}_n  \cdot  {\bf{p}}_n  } \over{E_n}} 
\delta^3 ( {\bf{x}} - {\bf{x}}_n(t) ) 
\EE
Eq.(\ref{lagrangian1}) is finally replaced by
\BE
\label{lagrangian}
 {\cal L}_{\rm eff} (\tilde{\sigma})=  
{{ \langle\Phi\rangle^2}\over{2}} \tilde{\sigma}
       [ c^2_s \Delta - {{\partial^2 } \over{\partial t^2}}] \tilde{\sigma}
      - \tilde{\sigma}  T^{\mu}_{\mu} 
\EE
In this way, we get the equation of motion
\BE
\label{step1}
       [ c^2_s \Delta - {{\partial^2 } \over{\partial t^2}}] \tilde{\sigma}=
         { { T^{\mu}_{\mu} }\over{ \langle \Phi \rangle^2}}
\EE
and it is clear that, for those
very large values $c_s\to \infty $ suggested by the properties of the vacuum, 
the $\tilde{\sigma}$ Green's function has practically no retardation effects. 
In this limit, Eq.(\ref{step1}) reduces to an instantaneous interaction
\BE
\label{step2}
          \Delta \tilde{\sigma}= 
{{T^{\mu}_{\mu}}\over{c^2_s \langle\Phi\rangle^2}} 
\EE
of vanishingly small strength. 
Finally for very slow motions, when the trace of the energy-momentum tensor 
reduces to the mass density
\BE
\label{source}
         \rho(x ) \equiv \sum_n m_n
\delta^3( {\bf{x}} - {\bf{x}}_n(t)) 
\EE
one gets, formally, a Poisson equation where, however, the Newton constant
$G_N=M^{-2}_{\rm Planck}$ is replaced by the product 
$c^{-2}_s \langle\Phi\rangle^{-2}$
\BE
\label{poisson}
            \Delta \tilde{\sigma}= 
{{1}\over{c^2_s \langle\Phi\rangle^2}}~ \rho( x)
\EE
For this reason, again, one may be tempted to exploit the identification
$M_{\rm Planck}=c_s \langle\Phi\rangle$ \cite{note1,note2}.

Before concluding, we comment on 
the momentum $\delta$ Eq.(\ref{beyond}) associated with the 
transition between the two branches of the
spectrum Eqs.(\ref{Eb}) and (\ref{Ea}). 
For $r \sim R_{\rm mfp} = \delta^{-1}$ the 
interparticle potential is not a simple $1/r$, as for asymptotic distances, 
but has to be computed from the 
Fourier transform of the $h-$field propagator
\BE
D(r)=
\int {{d^3 {\bf{p}} }\over{(2\pi)^3 }}
 {{e^{ i {\bf{p}}\cdot {\bf{r}} } }\over{ 
\tilde{E}^2({\bf{p}}) }}
\EE
and depends on the detailed form of the spectrum that interpolates between
Eqs.(\ref{Eb}) and (\ref{Ea}). However, 
in the Standard Model, for $M_h={\cal O}(\langle\Phi\rangle)$ and 
$c_s\langle\Phi\rangle = M_{\rm Planck}$, 
one would predict in any case 
a length scale $R_{\rm mfp}=\delta^{-1}$ in the millimeter
range. In this framework, 
the tight infrared-ultraviolet connection embodied in the relation 
$\delta \sim {{\langle\Phi\rangle^2}\over {M_{\rm Planck}}}$ 
would formally be identical to that occurring 
in models \cite{dimo} with extra space-time dimensions 
compactified at a size $R_{\rm c}=R_{\rm mfp}$.

\vskip 7pt
{\bf 5.}~Summarizing: 
the conventional description of the singlet Higgs field as a purely
massive field 
has to be modified to take into account that the energy spectrum has the 
form $c_s|{\bf{p}}|$ for ${\bf{p}} \to 0$.
In this sense, 
the Higgs condensate is a truly physical medium that can support
long-wavelength density fluctuations. However, 
their velocity
$c_s$ becomes infinitely large and their wavelengths become infinitely 
long when approaching the continuum limit. 
Physically, this corresponds to treat the Higgs condensate 
in a perfect-fluid approximation.

In a cutoff theory, $c_s$ is finite and this amounts
to introduce a finite mean free path $R_{\rm mfp} \equiv \delta^{-1}$ 
for the elementary constituents in the
condensate. In this situation, 
one finds a weak, attractive long-range force
proportional to ${{1}\over{c^2_s}}$ and
the energy spectrum acquires its single-particle form 
$\sqrt{ {\bf{p}}^2 + M^2_h}$ at very small momenta 
$|{\bf {p}}| \sim \delta$ with
$\delta= {\cal O}({{M_h}\over{c_s}})$. There are some
arguments for the identification 
$c_s \langle \Phi \rangle = M_{\rm Planck}$, that suggest
the possible 
relevance of our picture in connection with the problem of gravity 
\cite{note1,note2}. In this case, the length 
$R_{\rm mfp} $ would be a fundamental scale placed in the millimeter range.
\vskip 7 pt
{\bf Acknowledgements}~~~I thank P. M. Stevenson for 
useful discussions and collaboration.


\begin{thebibliography} {99}
\bibitem{note0}
The attractive tail originates from ultraviolet finite parts
of higher loop graphs \cite{mech} that cannot
be reabsorbed into a standard 
re-definition of the tree-level hard-core
potential. This effect is essential for a physical description of the
condensation process 
when approaching the limit of a vanishing renormalized mass
for the elementary quanta of the `empty', 
$\langle\Phi\rangle=0$ vacuum.
\end{thebibliography}
\end{document}